\documentclass[a4paper,12pt]{article}
\input{tcilatex}
\begin{document}

\setcounter{page}{0} \topmargin 0pt \oddsidemargin 5mm \renewcommand{%
\thefootnote}{\fnsymbol{footnote}} \newpage \setcounter{page}{0} 
\begin{titlepage}
\begin{flushright}
Berlin Sfb288 Preprint  \\
US-FT/05-00\\
hep-th/0004089\\
\end{flushright}
\vspace{0.2cm}
\begin{center}
{\Large {\bf Form Factors of the  Homogeneous
Sine-Gordon models} }

\vspace{0.8cm}
{\large  O.A.~Castro-Alvaredo$^\sharp$, A.~Fring$\,^\star$, C.~Korff$\,^\star$}

\vspace{0.2cm}
{\em $^\sharp$Departamento de F\'\i sica de Part\'\i culas, Facultad de F\'\i sica \\
Universidad de Santiago de Compostela\\
E-15706 Santiago de Compostela, Spain\\
\smallskip
$^\star$Institut f\"ur Theoretische Physik, 
Freie Universit\"at Berlin\\ 
Arnimallee 14, D-14195 Berlin, Germany }
\end{center}
\vspace{0.5cm}
 
\renewcommand{\thefootnote}{\arabic{footnote}}
\setcounter{footnote}{0}

\begin{abstract}
We provide general determinant  formulae for all n-particle form factors related to the
trace of the energy momentum tensor and the analogue of the order and disorder operator
in the $SU(3)_2$-homogeneous Sine-Gordon model. We employ the form factors related to
the trace of the energy momentum tensor in the application of the c-theorem and find perfect
agreement  with the physical picture recently obtained  by means of the thermodynamic 
Bethe ansatz. For finite resonance parameter we recover the expected WZNW-coset 
central charge and for infinite resonance parameter the theory decouples into two free
fermions.

\end{abstract}
\vfill{ \hspace*{-9mm}
\begin{tabular}{l}
\rule{6 cm}{0.05 mm}\\
Castro@fpaxp1.usc.es\\
Fring@physik.fu-berlin.de \\
Korff@physik.fu-berlin.de 
\end{tabular}}
\end{titlepage}
\newpage 

\section{Introduction}

In \cite{CFKM} a certain physical picture for the quantum field theory of
the Homogeneous Sine-Gordon models (HSG) \cite{HSG} was extracted from a
thermodynamic Bethe ansatz analysis. The central aim of this manuscript is
to inspect the picture for consistency by means of the form factor approach 
\cite{Kar,Smir}.

The HSG-models have been constructed as integrable perturbations of
WZNW-models. The related scattering matrices belong to a general class \cite
{HSGS,FK}, which describe the scattering of particles labeled by two quantum
numbers, where each of them may be associated to a simple Lie algebra.
Characteristic features of these S-matrices are the breaking of the parity
invariance of some amplitudes and in addition the presence of a resonance
parameter which enables the formation of unstable bound states. In \cite
{CFKM} we recovered the expected Virasoro coset central charge and found
that when the resonance parameter tends to infinity the system decouples
into several copies of minimal affine Toda field theories. Since the
ultraviolet central charge is also accessible by the c-theorem, the findings
in \cite{CFKM} may be checked for consistency.

Our manuscript is organised as follows: In section 2 we recall the general
properties of form factors. In section 3 we specialise the equations to the
case of the $SU(3)_{2}$-HSG model and provide the general solutions related
to the energy momentum tensor and the analogue of the order and disorder
operators. Our conclusions and a further outlook are presented in section 4.

\section{Generalities on Form Factors}

In order to fix our conventions and to set up the general framework we
commence by recalling briefly some general properties of form factors. For a
proper justification of them in terms of general principles of quantum field
theory and analytic properties in the complex plane we refer the reader to 
\cite{Kar,Smir,Zamocorr,BFKZ}.

Form factors are tensor valued functions, representing matrix elements of
some local operator $\mathcal{O}(\vec{x})$ at the origin between a
multiparticle in-state and the vacuum, which we denote by

\begin{equation}
F_{n}^{\mathcal{O}|\mu _{1}\ldots \mu _{n}}(\theta _{1},\ldots ,\theta
_{n}):=\left\langle 0|\mathcal{O}(0)|V_{\mu _{1}}(\theta _{1})V_{\mu
_{2}}(\theta _{2})\ldots V_{\mu _{n}}(\theta _{n})\right\rangle _{\text{in}%
}\,\,.
\end{equation}
Here the $V_{\mu }(\theta )$ are some vertex operators representing a
particle of species $\mu $ depending on the rapidity $\theta $ satisfying
the so-called Zamolodchikov algebra.

As a consequence of CPT-invariance or the braiding of two operators $V_{\mu
}(\theta )$ one obtains 
\begin{equation}
F_{n}^{\mathcal{O}|\ldots \mu _{i}\mu _{i+1}\ldots }(\ldots ,\theta
_{i},\theta _{i+1},\ldots )=F_{n}^{\mathcal{O}|\ldots \mu _{i+1}\mu
_{i}\ldots }(\ldots ,\theta _{i+1},\theta _{i},\ldots )S_{\mu _{i}\mu
_{i+1}}(\theta _{i,i+1})\,.  \label{W1}
\end{equation}
As usual we abbreviate $\theta _{ij}=$ $\theta _{i}-\theta _{j}$. The
analytic continuation in the complex $\theta $-plane at the cuts when $%
\theta =2\pi i$ together with crossing leads to 
\begin{equation}
F_{n}^{\mathcal{O}|\mu _{1}\ldots \mu _{n}}(\theta _{1}+2\pi i,\ldots
,\theta _{n})=F_{n}^{\mathcal{O}|\mu _{2}\ldots \mu _{n}\mu _{1}}(\theta
_{2},\ldots ,\theta _{n},\theta _{1})\,\,.  \label{W2}
\end{equation}
Since we are describing relativistically invariant theories we expect for an
operator $\mathcal{O}$ with spin $s$%
\begin{equation}
F_{n}^{\mathcal{O}|\mu _{1}\ldots \mu _{n}}(\theta _{1}+\Delta ,\ldots
,\theta _{n}+\Delta )=e^{s\Delta }F_{n}^{\mathcal{O}|\mu _{1}\ldots \mu
_{n}}(\theta _{1},\ldots ,\theta _{n})\,\,.  \label{rel}
\end{equation}
For a form factor whose first two particles are conjugate to each other we
have a kinematical pole at $i\pi $, which leads to a recursive equation
relating the (n-2)- and the n-particle form factor 
\begin{equation}
\stackunder{{\small \bar{\theta}}_{0}\rightarrow {\small \theta }_{0}}{\text{%
Res}}{\small F}_{n+2}^{\mathcal{O}|\bar{\mu}\mu \mu _{1}\ldots \mu _{n}}%
{\small (\bar{\theta}}_{0}{\small +i\pi ,\theta }_{0}{\small ,\theta }_{1}%
{\small ,\ldots ,\theta }_{n}{\small )}=i(1-\omega \prod_{l=1}^{n}S_{\mu \mu
_{l}}(\theta _{0l})){\small F}_{n}^{\mathcal{O}|\mu _{1}\ldots \mu _{n}}%
{\small (\theta }_{1}{\small ,\ldots ,\theta }_{n}{\small ),}  \label{kin}
\end{equation}
with $\omega $ being the factor of local commutativity and $\bar{\mu}$ the
anti-particle of $\mu $. We restrict our initial considerations to a model
in which stable bound states may not be formed and therefore we do not need
to report the so-called bound state residue equation.

To be able to associate a solution of the equations (\ref{W1})-(\ref{kin})
to a particular operator, the following upper bound on the asymptotic
behaviour \cite{DM} 
\begin{equation}
\left[ F_{n}^{\mathcal{O}|\mu _{1}\ldots \mu _{n}}(\theta _{1},\ldots
,\theta _{n})\right] _{i}\leq \,\Delta  \label{bound}
\end{equation}
turns out to be very useful. Here $\Delta $ denotes the conformal dimension
of the operator $\mathcal{O}$ in the conformal limit. For convenience we
introduced the short hand notation $\lim_{\theta _{i}\rightarrow \infty
}f(\theta _{1},\ldots ,\theta _{n})=$ const $\exp ([f(\theta _{1},\ldots
,\theta _{n})]_{i}\theta _{i})$.

Ultimately form factors serve to compute correlation functions, but they may
also be exploited to extract various other properties as for instance the
difference between the ultraviolet and infrared Virasoro central charges, as
stated in the so-called c-theorem \cite{ZamC} 
\begin{equation}
\Delta c=\sum_{n=1}^{\infty }\sum_{\mu _{1}\ldots \mu _{n}}\frac{9}{n!(2\pi
)^{n}}\int\limits_{-\infty }^{\infty }\ldots \int\limits_{-\infty }^{\infty }%
\frac{d\theta _{1}\ldots d\theta _{n}}{\left( \sum_{i=1}^{n}m_{\mu
_{i}}\cosh \theta _{i}\right) ^{4}}\left| F_{n}^{\mathcal{O}|\mu _{1}\ldots
\mu _{n}}(\theta _{1},\ldots ,\theta _{n})\right| ^{2}\,\,.  \label{cth}
\end{equation}
It is essentially the property (\ref{cth}) which we wish to employ for our
purposes and check for consistency of the physical picture which emerged in 
\cite{CFKM}.

\section{The SU(3)$_{2}$-HSG model}

For finite resonance parameter the SU(3)$_{2}$-HSG model describes the
WZNW-coset model with central charge $c=6/5$ perturbed by an operator with
conformal dimension $\Delta =3/5$. The model contains only two
self-conjugate solitons which are conveniently denoted by ``$+$'' and ``$-$%
'', since that will allow for compact notations. The S-matrix elements read 
\cite{HSGS} 
\begin{equation}
S_{\pm \pm }=-1\quad \quad \text{and}\quad \quad S_{\pm \mp }(\theta )=\pm
\tanh \frac{1}{2}\left( \theta \pm \sigma -i\frac{\pi }{2}\right) \;.
\label{ZamS}
\end{equation}
This means the scattering of particles of the same type is simply described
by the S-matrix of the thermal perturbation of the Ising model. Also the
remaining amplitudes do not possess poles inside the physical sheet, such
that the formation of stable particles via fusing is not possible. For
vanishing resonance parameter $\sigma $ the amplitudes $S_{\pm \mp }$
coincides formally with the ones which describe the massless flow between
the tricritical Ising and the critical Ising model as analysed in \cite{DMS}%
. However, there is an important conceptual difference since we view the
expressions (\ref{ZamS}) as describing the scattering of massive particles.
This has important consequences on the construction of the form factors and
in fact the solution we compute below will be different from the one
proposed in \cite{DMS}. In the HSG setting the massless flow was recovered
in the context of the thermodynamic Bethe ansatz \cite{CFKM} only as a
subsystem in terms of specially introduced variables combining the inverse
temperature and the resonance parameter. When the resonance parameter tends
to infinity the amplitudes $S_{\pm \mp }$ become one, describing
non-interacting scattering, such that the ``+''-system and the ``-''-system
decouple.

Attempting now to solve the equations presented in section 2, we proceed as
usual in this context and we make a factorization ansatz which already
extracts explicitly some of the singularity structure we expect to find. For
the case at hand we have to have a kinematical pole at $i\pi $ when two
particles are conjugate to each other 
\begin{equation}
F_{n}^{\mathcal{O}|\stackrel{l\,\times \,\pm }{\overbrace{\mu _{1}\ldots \mu
_{l}}}\stackrel{m\,\times \,\mp }{\overbrace{\mu _{l+1}\ldots \mu _{n}}}%
}(\theta _{1}\ldots \theta _{n})=H_{n}^{\mathcal{O}|\mu _{1}\ldots \mu
_{n}}Q_{n}^{\mathcal{O}|\mu _{1}\ldots \mu _{n}}(x_{1}\ldots
x_{n})\,\!\!\prod_{i<j}\!\!\frac{F_{\text{min}}^{\mu _{i}\mu _{j}}(\theta
_{ij})}{\left( x_{i}^{\mu _{i}}+x_{j}^{\mu _{j}}\right) ^{\delta _{\mu
_{i}\mu _{j}}}}.  \label{fact}
\end{equation}
We introduced the variable $x_{i}=\exp \theta _{i}$. The $H_{n}^{\mathcal{O}%
|\mu _{1}\ldots \mu _{n}}$ are normalization constants. As common we suppose
that the so-called minimal form factor satisfies 
\begin{equation}
F_{\text{min}}^{ij}(\theta )=F_{\text{min}}^{ji}(-\theta )S_{ij}(\theta )=F_{%
\text{min}}^{ji}(2\pi i-\theta )\,\,  \label{mini}
\end{equation}
and has neither zeros nor poles in the physical sheet. Then, if we further
assume that $Q_{n}^{\mathcal{O}|\mu _{1}\ldots \mu _{n}}(\theta _{1},\ldots
,\theta _{n})$ is separately symmetric in the first $l$ and the last $m$
rapidities and in addition 2$\pi i$-periodic function in all rapidities, the
ansatz (\ref{fact}) solves Watson's equations (\ref{W1}) and (\ref{W2}) by
construction. In particular we have 
\begin{equation}
Q_{n}^{\mathcal{O}|\stackrel{l\,\times \,+}{\overbrace{\mu _{1}\ldots \mu
_{l}}}\stackrel{m\,\times \,-}{\overbrace{\mu _{l+1}\ldots \mu _{n}}}%
}(x_{1},\ldots ,x_{n})=Q_{n}^{\mathcal{O}|\stackrel{m\,\times \,-}{%
\overbrace{\mu _{l+1}\ldots \mu _{n}}}\stackrel{l\,\times \,+}{\overbrace{%
\mu _{1}\ldots \mu _{l}}}}(x_{l+1},\ldots x_{n},x_{1},\ldots x_{l})\,\,,
\label{sym}
\end{equation}
such that when we have constructed a solution for one particular ordering of
the $\mu $'s, e.g. the upper sign in (\ref{fact}), we can obtain the
solution for a permuted ordering by the monodromy properties. Especially the
reversed order we obtain by applying equation (\ref{sym}). Despite the fact
that we do not gain anything new, it is still instructive to verify (\ref
{kin}) as a consistency check also for the different ordering. The monodromy
properties allow some simplification in the notation and from now on we
restrict our attention w.l.g. to the upper sign in (\ref{fact}). In addition
we deduce from equation (\ref{rel}) that for a spinless operator $\mathcal{O}
$ the total degree of $Q_{n}^{\mathcal{O}}$ has to be $l(l-1)/2$- $m(m-1)/2$.

A solution for the minimal form factors, i.e. of equations (\ref{mini}), is
found easily 
\begin{eqnarray}
F_{\text{min}}^{\pm \pm }(\theta ) &=&-i\sinh \frac{\theta }{2} \\
F_{\text{min}}^{\pm \mp }(\theta ) &=&\mathcal{N}^{\pm }(\theta
)\prod_{k=1}^{\infty }\tfrac{\Gamma (k+\frac{1}{4})^{2}\Gamma \left( k+\frac{%
1}{4}+\frac{i}{2\pi }(\theta \pm \sigma )\right) \Gamma \left( k-\frac{3}{4}-%
\frac{i}{2\pi }(\theta \pm \sigma )\right) }{\Gamma (k-\frac{1}{4}%
)^{2}\Gamma \left( k-\frac{1}{4}-\frac{i}{2\pi }(\theta \pm \sigma )\right)
\Gamma \left( k+\frac{3}{4}+\frac{i}{2\pi }(\theta \pm \sigma )\right) }\,\,
\\
&=&\mathcal{N}^{\pm }(\theta )\exp \left( -\int\limits_{0}^{\infty }\tfrac{dt%
}{t}\tfrac{\sin ^{2}\left( (i\pi -\theta \mp \sigma )\frac{t}{2\pi }\right) 
}{\sinh t\cosh t/2}\right) \,\,\,\,\,.  \label{14}
\end{eqnarray}
Here $F_{\text{min}}^{\pm \pm }(\theta )$ is the well-known minimal form
factor of the thermally perturbed Ising model \cite{BKW,YZam} and for the
upper choice of the signs, equation (\ref{14}) coincides for vanishing $%
\sigma $ up to normalization with the expression found in \cite{DMS}. We
introduced the normalization function $\mathcal{N}^{\pm }(\theta )=2^{\frac{1%
}{4}}\exp \left( \tfrac{i\pi (1\mp 1)\pm \theta }{4}-\tfrac{G}{\pi }\right) $
with $G=0.91597$ being the Catalan constant. The minimal form factors
possess various properties which we would like to employ in the course of
our argumentation. They obey the functional identities 
\begin{eqnarray}
F_{\text{min}}^{\pm \pm }(\theta +i\pi )F_{\text{min}}^{\pm \pm }(\theta )
&=&-\frac{i}{2}\sinh \theta  \label{he1} \\
F_{\text{min}}^{\pm \mp }(\theta +i\pi )F_{\text{min}}^{\pm \mp }(\theta )
&=&\frac{i^{\frac{2\mp 1}{2}}\,e^{\pm \frac{\theta }{2}}}{\cosh \frac{1}{2}%
\left( \theta \pm \sigma -\frac{i\,\pi }{2}\right) }\,\,.  \label{he2}
\end{eqnarray}
We will also exploit the asymptotic behaviour 
\begin{equation}
\lim\limits_{\sigma \rightarrow \infty }F_{\text{min}}^{\pm \mp }(\pm \theta
)\sim e^{-\frac{\sigma }{4}},\qquad \left[ F_{\text{min}}^{\pm \pm }(\theta
_{ij})\right] _{i}=\frac{1}{2},\qquad \left[ F_{\text{min}}^{\pm \mp
}(\theta _{ij})\right] _{i}=\QATOPD\{ . {0}{-1/2}\,\,.
\end{equation}
Together with the factorization ansatz (\ref{fact}) this leads us
immediately to the relations 
\begin{eqnarray}
\left[ F_{n}^{\mathcal{O}|l,m}\right] _{i} &=&\left[ Q_{n}^{\mathcal{O}%
|l,m}\right] _{i}+\frac{1-l}{2}\qquad \qquad \,\,\,\text{for\quad }1\leq
i\leq l  \label{as1} \\
\left[ F_{n}^{\mathcal{O}|l,m}\right] _{i} &=&\left[ Q_{n}^{\mathcal{O}%
|l,m}\right] _{i}+\frac{m-l-1}{2}\,\,\quad \quad \text{for \quad }l<i\leq
n\,,  \label{as2}
\end{eqnarray}
which are useful in the identification process of a particular solution with
a specific operator. Since we may restrict our attention to one particular
ordering only, we abbreviate the r.h.s. of (\ref{fact}) from now on as $%
F_{n}^{\mathcal{O}|l,m}$ and similar for the $Q$'s.

Substituting the ansatz (\ref{fact}) into the kinematic residue equation (%
\ref{kin}) reduces, with the help of (\ref{he1}) and (\ref{he2}), the whole
problem of determining the form factors to the following recursive equations 
\begin{eqnarray}
Q_{n+2}^{\mathcal{O}|l+2,m}(-x,x,\ldots ,x_{n}) &=&D_{n}^{l,m}(x_{1},\ldots
,x_{n})Q_{n}^{\mathcal{O}|l,m}(x_{1},\ldots ,x_{n})  \label{Qrec} \\
D_{n}^{l,m}(x,x_{1},\ldots ,x_{n}) &=&\frac{1}{2}(-ix)^{l+1}\sigma
_{l}^{+}\sum_{k=0}^{m}(-ie^{\sigma }x)^{-k}(1-\omega (-1)^{l+k})\sigma
_{k}^{-}
\end{eqnarray}
Here we introduced yet another short hand notation, namely for elementary
symmetric polynomials $\sigma _{k}(x_{1},\ldots ,x_{l})\equiv \sigma
_{k}^{+} $ and $\sigma _{k}(x_{l+1},\ldots ,x_{n})\equiv \sigma _{k}^{-}$ $%
\footnote{%
The elementary symmetric polynomials are generated by 
\[
\prod_{k=1}^{n}(x+x_{k})=\sum_{k=0}^{n}x^{n-k}\sigma _{k}(x_{1},\ldots
,x_{n})\,\,,\text{ i.e.\qquad \thinspace }\sigma _{k}(x_{1},\ldots ,x_{n})\,=%
\frac{1}{2\pi i}\oint \frac{dz}{z^{n-k+1}}\prod\limits_{k=1}^{n}(z+x_{k}) 
\]
(For more properties see e.g. \cite{Don}.)}$. Below we shall also employ $%
\sigma _{k}$ when the polynomials depend on all $n$ variables, $\bar{\sigma}%
_{k}$ when they depend on the $n$ inverse variables, i.e. $x_{i}^{-1}$and $%
\hat{\sigma}_{k}$ when they depend on the $n$ variables $x_{i}e^{-\sigma }$.

The recursive equations for the constants turn out to be 
\begin{equation}
H_{n+2}^{\mathcal{O}|l+2,m}=i^{m}2^{2l-m+1}e^{\sigma m/2}H_{n}^{\mathcal{O}%
|l,m}\,\,.  \label{Hrec}
\end{equation}
Fixing one of the lowest constants, the solutions to these equations read 
\begin{equation}
H^{\mathcal{O}|2s+t,m}=i^{sm}2^{s(2s-m-1+2t)}e^{sm\sigma /2}H^{\mathcal{O}%
|t,m},\qquad t=0,1\,.  \label{consol}
\end{equation}
For specific operators we will provide below the explicit expressions for
the $H^{\mathcal{O}|l,m}$. Notice that there is a certain ambiguity
contained in the equations (\ref{Hrec}), i.e. we can multiply $H_{n}^{%
\mathcal{O}|l,m}$ by $i^{2l}$, $i^{2l^{2}}$ or $(-1)^{l}$ and produce a new
solution. However, since in practical applications we are usually dealing
with the absolute values of the form factors, these ambiguities will turn
out to be irrelevant.

\subsection{Solutions}

Whenever we consider $F_{n}^{\mathcal{O}|l,m}$ with $l$ even for vanishing
resonance parameter $\sigma $, we can use the kinematic residue equation (%
\ref{kin}) $l/2$-times and finally construct $F_{n}^{\mathcal{O}|0,m}$,
which should correspond to a form factor of the thermally perturbed Ising
model. In other words in that case we can always use the well-known
solutions $Q_{n}^{\mathcal{O}|0,m}$ as the initial condition for the
recursive problem (\ref{Qrec}).

\subsubsection{The energy momentum tensor $\Theta $}

The only non-vanishing form factor of the energy momentum tensor in the
thermally perturbed Ising model is well know to be 
\begin{equation}
F_{2}^{\Theta }(\theta )=-2\pi im^{2}\sinh (\theta /2)\,\,.  \label{Issol}
\end{equation}
From this equation we deduce immediately that $[F_{n}^{\Theta |l,2}]_{i}=1/2$%
, which serves on the other hand to fix $[Q_{n}^{\Theta |l,2}]_{i}$ with the
help of (\ref{as1}) and (\ref{as2}). Recalling that the energy momentum
tensor is proportional to the perturbing field \cite{Cardy} and the fact
that the conformal dimension of the latter is $\Delta =3/5$ for the $%
SU(2)_{3}$-HSG model, the value $[F_{n}^{\Theta |l,2}]_{i}=1/2$ is
compatible with the bound (\ref{bound}). As a further consequence of (\ref
{Issol}), we deduce 
\begin{equation}
H^{\Theta |0,2}=2\pi m_{-}^{2}
\end{equation}
as the initial value for the computation of all higher constants in (\ref
{consol}). The distinction between $m_{-}$ and $m_{+}$ indicates that in
principle the mass scales could be very different as discussed in \cite{CFKM}%
. Notice that $H^{\Theta |0,0}$ is reached only formally, since the
kinematic residue equation does not connect to the vacuum expectation value.
The initial values for the recursive equations (\ref{Qrec}) are 
\begin{equation}
Q_{2}^{\Theta |0,2}=x_{1}^{-1}+x_{2}^{-1}\quad \quad \text{and\quad \quad }%
Q_{2t}^{\Theta |0,2t}=0\quad \text{for }t\geq 2\,\,.
\end{equation}
Taking now $\omega =1$, the solutions to (\ref{Qrec}), with the same
asymptotic behaviour as the energy momentum tensor in the thermally
perturbed Ising model, are computed to 
\begin{equation}
Q_{2s+2t}^{\Theta |2s,2t}=(-1)^{(s+1)t}e^{-t\sigma }\sigma _{1}\bar{\sigma}%
_{1}(\sigma _{2s}^{+})^{s-t}(\sigma _{2t}^{-})^{1-t}\det \mathcal{A}^{\Theta
}\mathcal{\,\,\qquad }\text{for }t\geq 1,s\geq 1,  \label{solu1}
\end{equation}
where $\mathcal{A}^{\Theta }$ is a ($t+s-2$)$\times $($t+s-2$)-matrix whose
entries are given by 
\begin{equation}
\mathcal{A}_{ij}^{\Theta }=\QATOPD\{ . {\sigma _{2(j-i)+1}^{+}\qquad \qquad
\qquad \quad \text{for\quad }1\leq i<t}{(-1)^{(j-i+t)}\hat{\sigma}%
_{2(j-i+t)-1}^{-}\qquad \quad \quad \quad \text{for\quad }t\leq i\leq
s+t-2}\,\,.
\end{equation}
Explicitly we have 
\begin{equation}
\mathcal{A}^{\Theta }=\left( 
\begin{array}{rrrrrr}
\sigma _{1}^{+} & \sigma _{3}^{+} & \sigma _{5}^{+} & \sigma _{7}^{+} & 
\cdots  & 0 \\ 
0 & \sigma _{1}^{+} & \sigma _{3}^{+} & \sigma _{5}^{+} & \cdots  & 0 \\ 
\vdots  & \vdots  & \vdots  & \vdots  & \ddots  & \vdots  \\ 
0 & 0 & 0 & 0 & \cdots  & \sigma _{2s-1}^{+} \\ 
-\hat{\sigma}_{1}^{-} & \hat{\sigma}_{3}^{-} & -\hat{\sigma}_{5}^{-} & \hat{%
\sigma}_{7}^{-} & \cdots  & 0 \\ 
0 & -\hat{\sigma}_{1}^{-} & \hat{\sigma}_{3}^{-} & -\hat{\sigma}_{5}^{-} & 
\cdots  & 0 \\ 
\vdots  & \vdots  & \vdots  & \vdots  & \ddots  & \vdots  \\ 
0 & 0 & 0 & 0 & \cdots  & (-1)^{t}\hat{\sigma}_{2t-1}^{-}
\end{array}
\right) \,\,.  \label{sss}
\end{equation}
One may easily  verify case-by-case that (\ref{solu1}) is a solution of (\ref
{Hrec})  to relatively high orders in $s$ and $t$. A general proof of this
result, which we present elsewhere \cite{prep},  can be obtained by
exploiting the fact that the determinant of $\mathcal{A}$  may also be
represented in terms contour integrals 
\begin{eqnarray}
\det \mathcal{A}^{\Theta }\mathcal{\,\,} &=&(-1)^{(s+1)t}\oint du_{1}\ldots
\oint du_{t-1}\oint dv_{1}\ldots \oint
dv_{s-1}\prod\limits_{i=1}^{2s}\prod\limits_{j=1}^{t-1}\frac{u_{j}+x_{i}}{%
u_{j}^{2s+2j-2}}  \label{ss} \\
&&\!\!\!\!\!\!\!\!\!\!\!\!\!\!\!\!\!\!\!\!\times
\prod\limits_{i=1+2s}^{2s+2t}\prod\limits_{j=1}^{s-1}\frac{v_{j}+\hat{x}_{i}%
}{v_{j}^{2t+2j-2}}\!\!\prod_{1\leq i<j\leq
t-1}(u_{j}^{2}-u_{i}^{2})\!\!\prod_{1\leq i<j\leq
s-1}(v_{j}^{2}-v_{i}^{2})\prod_{j=1}^{s-1}%
\prod_{i=1}^{t-1}(u_{i}^{2}+v_{j}^{2})\,.  \nonumber
\end{eqnarray}
In order to establish the equivalence between (\ref{sss}) and (\ref{ss}) we
simply use the integral representation for the symmetric polynomals as
stated in the footnote. The integrals in (\ref{ss}) are understood as $\oint
dz\equiv (2\pi i)^{-1}\oint_{|z|=\varrho }dz$ with $\varrho $ being an
arbitrary positive real number.

\noindent Assembling now all the quantities we obtain for instance 
\begin{equation}
F_{4}^{\Theta |++--}(\theta _{1},\theta _{2},\theta _{3},\theta _{4})=\frac{%
-\pi m_{-}^{2}e^{(\theta _{31}+\theta _{42})/2}(2+\sum_{i<j}\cosh (\theta
_{ij}))}{2\cosh (\theta _{12}/2)\cosh (\theta _{34}/2)}\prod_{i<j}F_{\text{%
min}}^{\mu _{i}\mu _{j}}(\theta _{ij})\,.
\end{equation}
Having computed all form factors for the energy momentum tensor we are in
the position to apply the c-theorem, i.e. we can in principle evaluate (\ref
{cth}). For finite values of $\sigma $ we obtain 
\begin{equation}
\Delta c^{(2)}=1,\qquad \Delta c^{(4)}=1.197...,\qquad \Delta
c^{(6)}=1.199\ldots \,,\quad \text{for }\sigma <\infty 
\end{equation}
where in the notation $\Delta c^{(n)}$, the superscript $n$ indicates the
upper limit in (\ref{cth}). Thus, the expected value of $c=6/5$ is well
reproduced. Apart from $\Delta c^{(2)}$, in which case the calculation can
be performed analytically, the integrals in (\ref{cth}) are computed
directly by a brute force Monte Carlo integration.

When the resonance parameter tends to infinity the system decouples and we
are left with two non-interacting free fermions, such that the only
contribution in the sum (\ref{cth}) is twice the free fermion two-particle
contribution, such that 
\begin{equation}
\lim_{\sigma \rightarrow \infty }\Delta c=1\,\,.  \label{cinf}
\end{equation}
In order to see this we collect the leading order behaviours form our
general solution 
\begin{equation}
\lim_{\sigma \rightarrow \infty }H_{2s+2t}^{\Theta |2s,2t}\sim e^{st\sigma
},\quad \lim_{\sigma \rightarrow \infty }Q_{2s+2t}^{\Theta |2s,2t}\sim
e^{-(t+s-1)\sigma },\quad \lim_{\sigma \rightarrow \infty }\prod_{i<j}F_{%
\text{min}}^{\mu _{i}\mu _{j}}(\theta _{ij})\sim e^{-st\sigma },
\end{equation}
which means 
\begin{equation}
\lim_{\sigma \rightarrow \infty }F_{2s+2t}^{\Theta |2s,2t}\sim
e^{-(t+s-1)\sigma }\,\,.
\end{equation}
Hence the only non-vanishing form factors in this limit are $F_{2}^{\Theta
|0,2}$ and $F_{2}^{\Theta |2,0}$, which establishes (\ref{cinf}).

\subsubsection{The order operator $\Sigma $}

For the other sectors we may proceed similarly, i.e. viewing always the
thermally perturbed Ising model as a benchmark. Taking now $\omega =1$, we
recall the solution for the order operator 
\begin{equation}
F_{2s+1}^{\Sigma }(\theta _{1},\ldots ,\theta _{2s+1})=i^{s}F_{1}^{\Sigma
}\prod_{i<j}\tanh \tfrac{\theta _{ij}}{2}=i^{s}(2i)^{2s^{2}+s}F_{1}^{\Sigma
}\left( \sigma _{2s+1}\right) ^{s}\prod_{i<j}\tfrac{F_{\text{min}}^{\pm \pm
}(\theta _{ij})}{x_{i}+x_{j}}.  \label{solord}
\end{equation}
With this information we may fix the initial values of the recursive
equations (\ref{Qrec}) and (\ref{Hrec}) at once to 
\begin{equation}
Q_{2t+1}^{\Sigma |0,2t+1}=\left( \sigma _{2t+1}\right) ^{-t}=\left( \bar{%
\sigma}_{2t+1}\right) ^{t}\qquad \text{and\qquad }H^{\Sigma
|0,1}=F_{1}^{\Sigma }\,\,.
\end{equation}
Furthermore, we deduce from equation (\ref{solord}) that $[F_{n}^{\Sigma
|2s,2t+1}]_{i}=0$. Respecting these constraints we find as explicit
solutions 
\begin{equation}
Q_{2s+2t+1}^{\Sigma |2s,2t+1}=(-1)^{(s+1)t}(\sigma _{1})^{\frac{1}{2}%
}(\sigma _{2s}^{+})^{s-t-1}(\sigma _{1}^{-})^{-\frac{1}{2}}(\sigma
_{2t+1}^{-})^{-t}\det \mathcal{A}^{\Sigma }\mathcal{\,\,},  \label{sol2}
\end{equation}
where $\mathcal{A}^{\Sigma }\mathcal{\,}$is a ($t+s$)$\times $($t+s$)-matrix
whose entries are given by 
\begin{equation}
\mathcal{A}_{ij}^{\Sigma }\mathcal{\,}=\QATOPD\{ . {\sigma
_{2(j-i)}^{+}\qquad \qquad \qquad \qquad \quad \,\,\,\,\quad \text{for\quad }%
1\leq i\leq t}{(-1)^{(j-i+t+1)}\hat{\sigma}_{2(j-i+t)+1}^{-}\qquad \quad
\quad \quad \text{for\quad }t<i\leq s+t}\,\,.
\end{equation}
Explicitly this reads 
\begin{equation}
\mathcal{A}^{\Sigma }\mathcal{\,}=\left( 
\begin{array}{rrrrrr}
1 & \sigma _{2}^{+} & \sigma _{4}^{+} & \sigma _{6}^{+} & \cdots  & 0 \\ 
0 & 1 & \sigma _{2}^{+} & \sigma _{4}^{+} & \cdots  & 0 \\ 
\vdots  & \vdots  & \vdots  & \vdots  & \ddots  & \vdots  \\ 
0 & 0 & 0 & 0 & \cdots  & \sigma _{2s}^{+} \\ 
-\hat{\sigma}_{1}^{-} & \hat{\sigma}_{3}^{-} & -\hat{\sigma}_{5}^{-} & \hat{%
\sigma}_{7}^{-} & \cdots  & 0 \\ 
0 & -\hat{\sigma}_{1}^{-} & \hat{\sigma}_{3}^{-} & -\hat{\sigma}_{5}^{-} & 
\cdots  & 0 \\ 
\vdots  & \vdots  & \vdots  & \vdots  & \ddots  & \vdots  \\ 
0 & 0 & 0 & 0 & \cdots  & (-1)^{(t+1)}\hat{\sigma}_{2t+1}^{-}
\end{array}
\right) \,\,.
\end{equation}
Once again the determinant of $\mathcal{A}$ admits an integral representation
\begin{eqnarray}
\det \mathcal{A}^{\Sigma }\mathcal{\,\,} &=&(-1)^{s(t-1)}\oint du_{1}\ldots
\oint du_{t}\oint dv_{1}\ldots \oint
dv_{s}\prod\limits_{i=1}^{2s}\prod\limits_{j=1}^{t}\frac{u_{j}+x_{i}}{%
u_{j}^{2s+2j-1}} \\
&&\!\!\!\!\!\!\!\!\!\!\!\!\!\!\!\!\!\!\!\!\times
\prod\limits_{i=1+2s}^{2s+2t+1}\prod\limits_{j=1}^{s}\frac{v_{j}+\hat{x}_{i}%
}{v_{j}^{2t+2j-2}}\!\!\prod_{1\leq i<j\leq
t}(u_{j}^{2}-u_{i}^{2})\!\!\prod_{1\leq i<j\leq
s}(v_{j}^{2}-v_{i}^{2})\prod_{j=1}^{s}\prod_{i=1}^{t}(u_{i}^{2}+v_{j}^{2})\,
\nonumber
\end{eqnarray}
which may be used for a general proof  \cite{prep}.

When the resonance parameter tends to infinity we obtain the following
asymptotic behaviour 
\begin{eqnarray}
\quad \lim_{\sigma \rightarrow \infty }Q_{2s+2t+1}^{\mu |2s,2t+1} &\sim
&e^{-s\sigma } \\
\lim_{\sigma \rightarrow \infty }H_{2s+2t+1}^{\mu |2s,2t+1}\prod_{i<j}F_{%
\text{min}}^{\mu _{i}\mu _{j}}(\theta _{ij}) &=&\text{const\negthinspace
\negthinspace \negthinspace \negthinspace }\prod_{1\leq i<j\leq 2s}\text{%
\negthinspace \negthinspace \negthinspace \negthinspace }F_{\text{min}%
}^{++}(\theta _{ij})\text{\negthinspace \negthinspace \negthinspace
\negthinspace }\prod_{2s<i<j\leq 2s+2t+1}\text{\negthinspace \negthinspace
\negthinspace \negthinspace }F_{\text{min}}^{--}(\theta _{ij})\,.
\end{eqnarray}
This means unless $s=0$, that is a reduction to the thermally perturbed
Ising model, the form factors will vanish in this limit.

\subsubsection{The disorder operator $\mu $}

For the disorder operator we have $\omega =-1$ and the solution acquires the
same form as in the previous case 
\begin{equation}
F_{2s}^{\mu }(\theta _{1},\ldots ,\theta _{2s})=i^{s}F_{0}^{\mu
}\prod_{i<j}\tanh \frac{\theta _{ij}}{2}\,\,.
\end{equation}
Similar as for the order variable we can fix the initial values of the
recursive equations (\ref{Qrec}) and (\ref{Hrec}) to 
\begin{equation}
Q_{2t}^{\mu |0,2t}=\left( \sigma _{2t}\right) ^{1/2-t}=\left( \bar{\sigma}%
_{2t}\right) ^{t-1/2}\qquad \text{and\qquad }H^{\mu |0,0}=F_{0}^{\mu }\,\,.
\end{equation}
Furthermore, we deduce $[F_{n}^{\mu |2s,2t}]_{i}=0$. Respecting these
constraints we find as a general solution 
\begin{equation}
Q_{2s+2t}^{\mu |2s,2t}=(-1)^{st}(\sigma _{2s+2t})^{\frac{3}{2}-t}(\sigma
_{2s}^{+})^{s-2}(\sigma _{2t}^{-})^{-1}\det \mathcal{A}^{\mu }\mathcal{\,},
\label{sol3}
\end{equation}
where $\mathcal{A}^{\mu }$ is a ($t+s$)$\times $($t+s$)-matrix whose entries
are given by 
\begin{equation}
\mathcal{A}_{ij}^{\mu }=\QATOPD\{ . {\sigma _{2(j-i)}^{+}\qquad \qquad
\qquad \quad \,\,\,\,\quad \text{for\quad }1\leq i\leq t}{(-1)^{(j-i+t)}\hat{%
\sigma}_{2(j-i+t)}^{-}\qquad \quad \quad \quad \text{for\quad }t<i\leq
s+t}\,\,.
\end{equation}
Explicitly we have 
\begin{equation}
\mathcal{A}^{\mu }=\left( 
\begin{array}{rrrrrr}
1 & \sigma _{2}^{+} & \sigma _{4}^{+} & \sigma _{6}^{+} & \cdots  & 0 \\ 
0 & 1 & \sigma _{2}^{+} & \sigma _{4}^{+} & \cdots  & 0 \\ 
\vdots  & \vdots  & \vdots  & \vdots  & \ddots  & \vdots  \\ 
0 & 0 & 0 & 0 & \cdots  & \sigma _{2s}^{+} \\ 
1 & -\hat{\sigma}_{2}^{-} & \hat{\sigma}_{4}^{-} & -\hat{\sigma}_{6}^{-} & 
\cdots  & 0 \\ 
0 & 1 & -\hat{\sigma}_{2}^{-} & \hat{\sigma}_{4}^{-} & \cdots  & 0 \\ 
\vdots  & \vdots  & \vdots  & \vdots  & \ddots  & \vdots  \\ 
0 & 0 & 0 & 0 & \cdots  & (-1)^{t}\hat{\sigma}_{2t}^{-}
\end{array}
\right) \,\,.
\end{equation}
Similarly as in the previous sections we can write the determinant of $%
\mathcal{A}$ alternatively in form of an integral representation
\begin{eqnarray}
\det \mathcal{A}^{\mu }\mathcal{\,\,} &=&(-1)^{s(t-1)}\oint du_{1}\ldots
\oint du_{t}\oint dv_{1}\ldots \oint
dv_{s}\prod\limits_{i=1}^{2s}\prod\limits_{j=1}^{t}\frac{u_{j}+x_{i}}{%
u_{j}^{2s+2j-1}} \\
&&\!\!\!\!\!\!\!\!\!\!\!\!\!\!\!\!\!\!\!\!\times
\prod\limits_{i=1+2s}^{2s+2t}\prod\limits_{j=1}^{s}\frac{v_{j}+\hat{x}_{i}}{%
v_{j}^{2t+2j-1}}\!\!\prod_{1\leq i<j\leq
t}(u_{j}^{2}-u_{i}^{2})\!\!\prod_{1\leq i<j\leq
s}(v_{j}^{2}-v_{i}^{2})\prod_{j=1}^{s}\prod_{i=1}^{t}(u_{i}^{2}+v_{j}^{2})\,.
\nonumber
\end{eqnarray}

When the resonance parameter tends to infinity we observe the following
asymptotic behaviour 
\begin{eqnarray}
\quad \lim_{\sigma \rightarrow \infty }Q_{2s+2t}^{\mu |2s,2t}
&=&(-1)^{st}Q_{2s}^{\mu |2s,0}Q_{2t}^{\mu |0,2t}\quad \\
\lim_{\sigma \rightarrow \infty }H_{2s+2t}^{\mu |2s,2t}\prod_{i<j}F_{\text{%
min}}^{\mu _{i}\mu _{j}}(\theta _{ij}) &=&\text{const\negthinspace
\negthinspace \negthinspace }\prod_{1\leq i<j\leq 2s}\text{\negthinspace
\negthinspace \negthinspace }F_{\text{min}}^{++}(\theta _{ij})\text{%
\negthinspace \negthinspace \negthinspace }\prod_{2s<i<j\leq 2t+2s}\text{%
\negthinspace \negthinspace \negthinspace }F_{\text{min}}^{--}(\theta _{ij})
\end{eqnarray}
such that 
\begin{equation}
\lim_{\sigma \rightarrow \infty }F_{2s+2t}^{\mu |2s,2t}\sim F_{2t}^{\mu
|0,2t}F_{2s}^{\mu |2s,0}\,\,.
\end{equation}
This means also in this sector we observe the decoupling of the theory into
two free fermions.

\section{Conclusions}

The application of the c-theorem confirms very well the physical picture we
found in \cite{CFKM} from the thermodynamic Bethe ansatz. For finite
resonance parameter we recover the expected Virasoro central charge of $c=6/5
$ and for $\sigma \rightarrow \infty $ the theory decouples in all sectors
into two non-interacting free fermions. Besides the construction of all
n-particle form factors related to the trace of energy momentum, we computed
in addition the complete solutions for the order and disorder operator in
form of determinants whose entries are symmetric polynomials. Such
determinant formulae have occurred before in various places in the
literature, e.g. \cite{Zamocorr,deter}. Representing the solutions for form
factors in this form has turned out to be useful in the construction of
correlation functions \cite{det} and might eventually lead to a
reformulation of the whole problem in terms of differential equations
analogous to the situation in conformal field theory \cite{BPZ}. Apart from
higher spin solutions which may always be constructed by including the
polynomials as suggested in \cite{CardyM}, we did not find any additional
solutions related to other sectors. We expect that a careful analysis of the
cluster decomposition property will lead to more conclusive statements
concerning the question whether such solutions exist at all. From a
mathematical point of view it is also desirable to present a rigorous proof
of the determinant formulae \cite{prep}.

\noindent \textbf{Acknowledgments: } A.F. and C.K. are grateful to the
Deutsche Forschungsgemeinschaft (Sfb288) for financial support. O.A.C.
thanks CICYT (AEN99-0589), DGICYT (PB96-0960), and the EC Commission (TMR
grant FMRX-CT96-0012) for partial financial support and is also very
grateful to the Institut f\"{u}r theoretische Physik of the Freie
Universit\"{a}t for hospitality. We are grateful to J.L. Miramontes and G.
Mussardo for useful discussions.

\begin{description}
\item  {\small \setlength{\baselineskip}{12pt}}
\end{description}

\end{document}